\documentclass[prb,preprint,superscriptaddress]{revtex4-1} 

\usepackage[english]{babel}
\usepackage[utf8x]{inputenc}

\usepackage{amsmath}  
\usepackage{amsfonts} 
\usepackage{graphicx} 
\usepackage{ifthen}
\usepackage{multirow}
\usepackage[dvipsnames]{xcolor}
\usepackage{subcaption}
\usepackage{titlesec}
\usepackage{makecell}
\usepackage{comment}

\newenvironment{iquote}
    {\vspace{.4\baselineskip}\itshape\list{}{\leftmargin=0.15in\rightmargin=0.15in}%
    \item\relax}
    {\endlist\vspace{.4\baselineskip}}

\newboolean{redactswitch}
\setboolean{redactswitch}{false} 

\newcommand{\redact}[1]{\ifthenelse{
    \boolean{redactswitch}}{{[Redacted]}}{
    {#1}}}

\newcommand{\fracroot}[2]{\ifthenelse{#1=1}{\frac{1}{\sqrt{#2}}}{\sqrt{\frac{#1}{#2}}}}

\newcommand{\who}[1]{\textbf{#1\emph{:}}}

\newboolean{showtimestamp}
\setboolean{showtimestamp}{false}

\newcommand{\timestamp}[1]{
    \ifthenelse{\boolean{showtimestamp}}{\textcolor{red}{Timestamp: #1}}{}}

\usepackage{hyperref}  
\hypersetup{colorlinks=true,urlcolor=blue,citecolor=blue,linkcolor=blue}   
\usepackage{enumitem}        
\setlist{nosep}

\usepackage{setspace}

\begin{document}



\title{How media hype affects our physics teaching: \\ A case study on quantum computing}

\author{\redact{Josephine C.\ Meyer}}
\affiliation{\redact{Department of Physics, University of Colorado Boulder, Boulder, CO 80309}}
\author{\redact{Gina Passante}}
\affiliation{\redact{Department of Physics, California State University Fullerton, Fullerton, CA 92831}}
\author{\redact{Steven J.\ Pollock}}
\affiliation{\redact{Department of Physics, University of Colorado Boulder, Boulder, CO 80309}}
\author{\redact{Bethany R.\ Wilcox}}
\affiliation{\redact{Department of Physics, University of Colorado Boulder, Boulder, CO 80309}}


\date{\today}



\maketitle 

Popular media is an unspoken yet ever-present element of the physics landscape and a tool we can utilize in our teaching.\cite{Shipman:2000} It is also well-understood that students enter the physics classroom with a host of conceptions learned from the world at large.\cite{diSessa:2000, Redish:2003} It stands to reason, then, to suspect that media coverage
may be a major contributing factor to students' views on physical phenomena and the nature of science\cite{studentanecdote} -- one whose influence will only grow amid the 21st century digital age. Yet the role of the media in shaping physics teaching and learning has remained largely unexplored in the physics education research (PER) literature so far. 


Here, we explore the phenomenon of media hype from a theoretical and practical perspective: how media rhetoric of current topics in science and technology evolves, and how it affects students and instructors. We argue that media hype of cutting-edge science can be a double-edged sword for educators, with the same amped-up rhetoric that motivates students to enter the classroom tending to result in inflated preconceptions of what the science and technology can actually do. We draw on examples related to teaching quantum computing as a case study, though the findings we present should generalize to other topics garnering significant media attention -- from exoplanets to graphene to batteries for electric vehicles. We conclude with a set of practical recommendations for physics teachers at all levels who wish to be more cognizant of the role exposure to popular media has on students and to tailor our teaching accordingly.

\section{Conceptualizing hype}
\label{sec:theory}

For our purposes, we conceptualize hype as optimistic coverage of an emerging technology designed to drum up excitement, usually for a non-expert audience. As its best, hype can be considered innocuous and even beneficial, often driven by researchers themselves seeking the attention of funders.\cite{Roberson:2021Minerva} At the other extreme is media coverage that dangerously misrepresents science in the public eye. It is thus useful to position hype on a spectrum, from uncertain but scientifically well-founded predictions about the future, to highly speculative claims based largely on conjecture, to the practically or scientifically absurd.\cite{footnote:examplesQC} 

\begin{figure}
    \centering
    \includegraphics[scale=0.5]{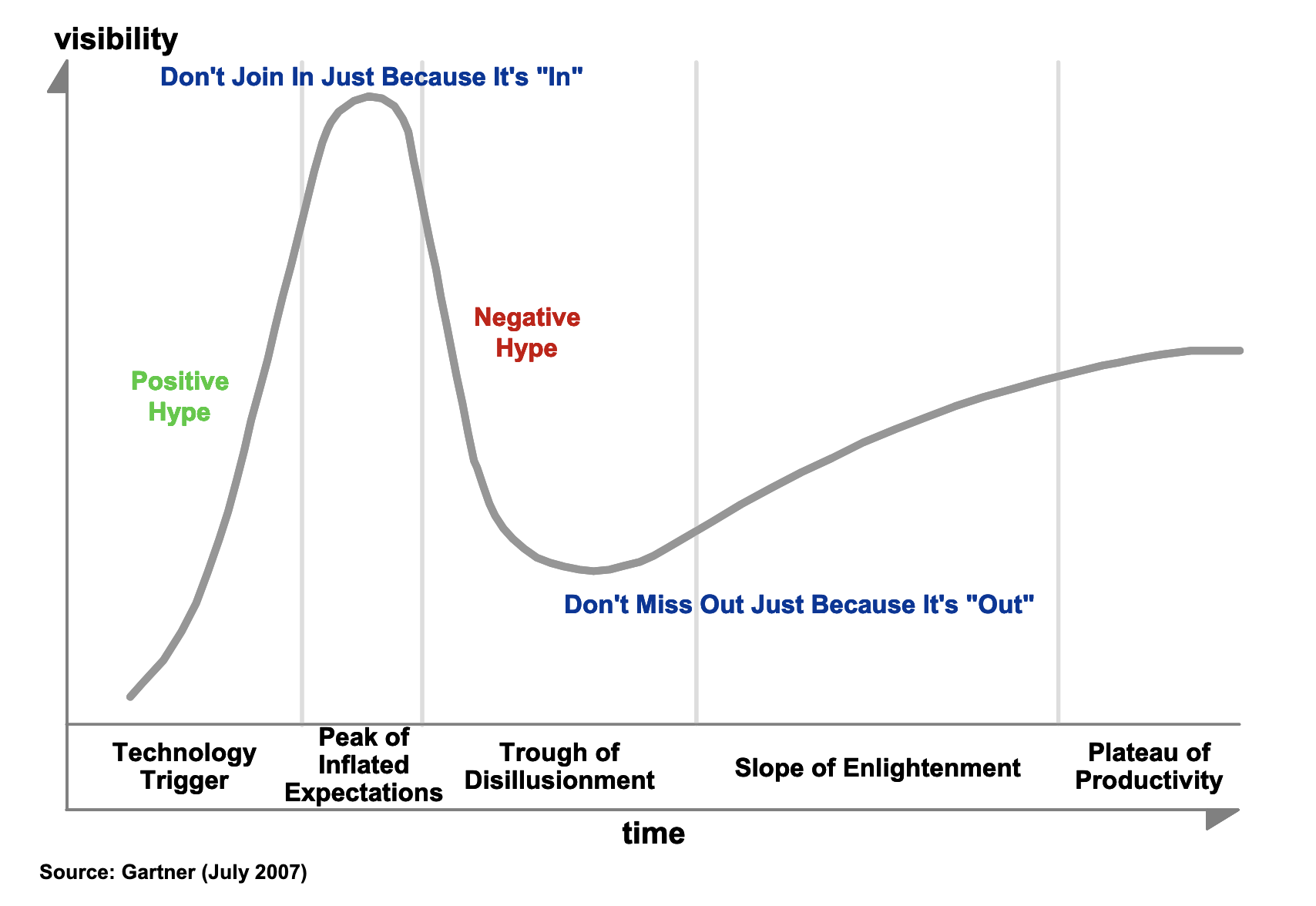}
    \caption{Gartner Hype Cycle. Source: Gartner, 2022.\cite{Gartner:2007}}
    \label{fig:gartnerhype}
\end{figure}

The Gartner Hype Cycle\cite{
Dedehayir:2016, Gartner:2018} from science, technology, and society studies is an empirical model for the effects of hype on emerging industries. As illustrated in Fig.~\ref{fig:gartnerhype}, the Gartner Hype Cycle posits that a technological innovation can spawn media attention above that warranted by the present merit of the technology. Eventually, expectations and publicity peak and then decline as impatience and setbacks deflate initial hopes. After reaching the so-called ``Trough of Disillusionment,'' attention begins to rise again once the technology matures enough to satisfy early adopters, finally plateauing at the technology's actual merit once adoption is widespread. Every year, Gartner Inc.\ publishes a list of new technologies along with their projected position along the Gartner Hype Cycle chart. Invariably, at least a few of these technologies will relate to topics we wish to teach in the physics classroom.

\section{The case of quantum computing} 

Quantum technology is among the areas of physics garnering extensive attention in the popular science media today. Controversial claims that we have reached ``quantum supremacy,''\cite{Arute:2019} that quantum technology will revolutionize machine learning,\cite{Musti:2020} or that China could use quantum computing to decode U.S.\ military secrets\cite{Titcomb:2021} have all made headlines in recent years. Our goal here is not to evaluate these claims, nor do we take a stance on whether such hype is fundamentally good or bad.\cite{Roberson:2020PUS} However, such claims have all provoked significant skepticism among experts and discussion is taking place in the community regarding quantum hype's effects on science and society.\cite{Smith:2020, Roberson:2021QST, Ezratty:2022} Indeed, quantum computing was classified near the foot of the ``Technology Trigger'' section in 2011\cite{Gartner:2011, Cuccureddu:2011} and as of 2021 is classified as ``Peak of Inflated Expectations''\cite{Gartner:2021} with a final plateau not anticipated for at least 10 years.

Our interest in the impact of media rhetoric on teaching began with a series of interviews with instructors teaching beyond-first-year undergraduate and hybrid undergraduate/graduate elective courses on quantum computing and associated topics, listed in physics and computer science departments.\cite{Meyer:2022PhysRev} Though our interview protocol was focused on instructor experiences teaching quantum computing and never specifically asked about the media or other outside influences on students, all six interviewees spontaneously mentioned media hype as a particularly salient influence on their teaching. We identify three important themes from these interviews that we expect can be broadly generalized across topics. The quotes we present were chosen for their clarity and representativeness. It is not our intention to validate or endorse specific instructor claims or teaching approaches, but rather simply to document the range of themes we heard from our interviewees demonstrating the effect of hype on their teaching and their diverse approaches in response.

If Gartner's analysis proves correct, the timeframe of this interview study (summer 2021) roughly coincides with the early stages of the peak of the quantum computing hype cycle, though since negative hype can be as narrative-dominating as positive hype we expect these themes to be applicable to technologies at any point in the hype cycle. 


\section{Findings: Teaching physics in an atmosphere of hype}

\subsection{Hype and student motivation}

All six instructors interviewed noted that student interest in their QIS courses is motivated in large part by popular media. In the words of instructor Franz:

\begin{iquote}
\who{Franz}
Why are they taking this elective instead of others? 
I think it's because there's so much hype ... 
They hear a lot about it in the media, and they perceive it to be moving quickly.
\timestamp{6:35 verified}
\end{iquote}

Edwin stated that, in his experience, popular science media such as Scientific American are a particularly important influence on prospective science students. Edwin -- who began teaching quantum computing in 2008 -- relates that media coverage was a major motivator for students even before the spike in popular attention. In his words:

\begin{iquote}
\who{Edwin}
Students are following a more refined version of popular media. 
So whether or not it was in USA Today, it was in the science sections they were looking at.
\timestamp{28:15 verified}
\end{iquote}

The elective courses in our study are strictly optional for students, so the effect of hype in students' decision to enroll will be particularly visible. However, generating and sustaining student motivation is essential for effective learning in virtually any physics setting. Our findings, thus, suggest that a brief unit on a topic garnering media attention could be a significant motivator for engagement in physics courses generally.

\subsection{Hype, prior knowledge, and myth-busting}

PER literature has long established that students do not enter the physics classroom as ''blank slates'' but rather with a host of preconceptions that instructors must help students refine. However, until now, the role of the media in forming these preconceptions has been largely overlooked in the literature. We heard consistently from instructors that though media attention gets students in the door, instructors observe that hype can also cause distorted preconceptions about quantum technology. In response, interviewees Albert, Carl, and Edwin
all explicitly design their courses with media coverage in mind.

For instance, Edwin finds it necessary to dispel myths in order to refine students' understandings of what quantum computing can do:

\begin{iquote}
\who{Edwin}
It's ... hype they've heard from the popular media ... that there's this completely novel notion of computation that involves ... 
magically follow[ing] every path simultaneously. And -- and that's false, of course ... I spent a lot of time explaining, you know, how that's an interpretation but not a particularly useful one.
\timestamp{13:32 verified}
\end{iquote}



Carl related the following as one of his primary goals for teaching his undergraduate quantum computing course: 

\begin{iquote}
\who{Carl}
My hope is that if we have a class like this ... 
it's contributing really to ... a somewhat more savvy and educated [workforce]. 
So that if someone comes along and tries to sell them snake oil, they'll say, ``Wait, I've studied, what you're saying is wrong ... you're kind of scamming us!'' 
\timestamp{25:20 verified}
\end{iquote}

Physics education research has shown the merit of the ``elicit, confront, resolve''\cite{Kryjevskaia:2014} approach in moving beyond students' naive preconceptions in mechanics courses. 
Carl and Edwin appear to be using an analogous technique with regard to students' preconceptions from media exposure, eliciting and then critically engaging with claims that students may have seen outside the classroom.
We believe this is a powerful approach for instructors to consider in all physics contexts, engaging with media claims as a way to disentangle truth from marketing.

\subsection{Hype affects content coverage}

\textit{This section is more technical and primarily of relevance to quantum computing instructors.}

In Sec.~\ref{sec:theory}, we argue for viewing hype on a spectrum, from innocuous and even beneficial to recklessly disconnected from the science. We find that this interpretation of hype mirrors real-world decisions instructors make as to what content to cover. Here, we focus on quantum algorithms, one of the few semi-universal concepts in quantum computation\cite{Seegerer:2021} and a key driver of quantum hype. (A quantum computer is functionally useless without practical algorithms to run on it, and algorithms that make headlines may be impractical in practice.)

In our interviews, instructors placed quantum algorithms themselves on a scale from proven and theoretically practical to patently absurd. At one end of the scale is Shor's integer-factoring algorithm,\cite{Shor:1994} which was the only algorithm taught by all six instructors and what instructor Carl called the ``one good algorithm'' for quantum computers. Shor's algorithm is unique in that it both solves a problem of real-world significance (one that could decrypt existing internet security protocols) and does so with an exponential speedup over classical computers. At the other end of the spectrum, some would argue, lie quantum algorithms such as for machine learning, that in Carl's words have been published for ``marketing purposes'' but have no proven speed-up over classical machines:

\begin{iquote}
\who{Carl}
I don't cover [such algorithms] for the same reason I don't cover, like, cold fusion and homeopathy!
\timestamp{1:11:00 verified}
\end{iquote}

In between these two extremes lie a variety of algorithms that are theoretically interesting but have limited practical value. Some such algorithms solve only contrived problems, while others produce a speedup too marginal to justify developing a quantum computer. Nonetheless, we heard from instructors that these algorithms can still be valuable pedagogically, for instance for the insight they provide into quantum theory or computer science theory. Franz would place Grover's search algorithm\cite{Grover:1996} -- arguably the second most famous quantum algorithm after Shor's -- in this category:

\begin{iquote}
\who{Franz}
Concerning hype, I'm not sure [Grover's algorithm] has any value in the world ... I still teach it because it is beautiful.
\end{iquote}

Designing a quantum computing course in the context of media hype involves pedagogical trade-offs. How does one design a course about an up-and-coming technology where there is really only ``one good algorithm'' worth running on a quantum computer? Are beautiful algorithms with no practical purpose worth spending class time on, or is it better to skip these topics completely, as instructor Albert does, to focus on the fundamental quantum mechanics? Or even, as Carl implies, are algorithms with little practical use actually pedagogically necessary to cover so that students can read quantum computing literature with an appropriate level of skepticism? These questions are worth pondering because they highlight the fundamental trade-offs that come with teaching hyped topics: what comes first, the technology or the physics? 

\section{Conclusion: Teaching physics in the context of media hype}

While media hype is inescapable for some instructors, all students enter our courses with conceptions driven consciously or subconsciously by media. When something like quantum computing makes headlines, it can be both a hook for getting students in the door and a challenge for instructors who must address misinformed preconceptions rooted in media rhetoric. The same is true of many other physics topics that garner significant media attention -- from the Higgs Boson to high-temperature superconductors.

As evidenced by the myriad approaches taken by the instructors we interviewed, there is no one right way to address media hype in the classroom. Moreover, media hype presumably has very different effects on students at different points in the hype cycle; instructors teaching a topic currently in the ``Trough of Disillusionment'' may find the need to challenge deflated perceptions as important as our interviewees found the need to temper inflated ones. The least we can do is familiarize ourselves with the recent mainstream and popular science media coverage and critically question whether the coverage is truly grounded in science. 

Finally, drawing on common wisdom and the demonstrated importance of media hype to our interviewees, we propose a set of questions that instructors may find valuable when teaching topics garnering significant media attention:

\begin{itemize}
    \item What is my goal in bringing the topic into the classroom? Is it to engage with the technology directly or to generate excitement about and appreciation for the underlying physics?
    \item What media coverage are my students likely to have been exposed to coming into my course? How scientifically accurate are the media portrayals of this technology?
    \item What preconceptions might students bring into the course as a result of media coverage? How might these ideas affect student engagement and learning?
    \item Where is the technology right now in the hype cycle?
    \item How can I leverage discussions of hyped technologies to help students become scientifically-literate citizens? Can I use these discussions to build skills such as critical thinking, ethical reasoning, \cite{Meyer:2022ASEE} or science communication \cite{Spektor-Levy:2009}?
    
\end{itemize}

\begin{acknowledgments}
\redact{Special thanks to our interviewees. This research is funded by the CU Boulder Department of Physics, the NSF GRFP, and NSF Grants No.'s 2012147 and 2011958.}

\end{acknowledgments}


\begin{thebibliography}{99}
\bibitem{studentanecdote}
\redact{JCM} relates an amusing anecdote in which an undergraduate who had just completed introductory mechanics attempted to explain a popular science article to her. She remarked that the student's interpretation of the article could not possibly be correct because it violated conservation of energy. Rather than admitting he may have misread the article, the student walked away from the conversation adamant that the laws of mechanics had been disproved!
\bibitem{footnote:examplesQC}
In the context of quantum computing, a statement in the relatively benign first category might be ''Shor's integer-factoring algorithm will break RSA encryption,'' a speculative statement in the second category might be ''quantum computing will revolutionize drug development,'' and a scientifically absurd claim we have heard far too often would be that ``quantum computers will allow us to solve NP-hard problems.''
\end{thebibliography}
\end{document}